\documentclass[prl,twocolumn,amsmath,amssymb,floatfix,showpacs,superscriptaddress]{revtex4}
\usepackage{graphicx}
\usepackage{amsmath}
\usepackage{bm}

\newcommand{\bq}{\begin{equation}}
\newcommand{\ee}{\end{equation}}
\newcommand{\fr}[2]{\frac{#1}{#2}}
\newcommand{\eps}{\varepsilon}

\renewcommand{\vec}[1]{\mathbf{#1}}
\usepackage{colordvi}
\usepackage{graphicx}
\usepackage{color}

\begin{document}
\title{Wigner crystal phases in bilayer graphene}
\date{\today }

\author{P. G. Silvestrov}
\affiliation{Institut f\"ur Mathematische Physik, TU Braunschweig,
Mendelssohnstr. 3, 38106 Braunschweig, Germany}
\author{P. Recher}
\affiliation{Institut f\"ur Mathematische Physik, TU Braunschweig,
Mendelssohnstr. 3, 38106 Braunschweig, Germany}
\affiliation{Laboratory for Emerging Nanometrology Braunschweig,
38106 Braunschweig, Germany}
\date{\today}

\begin{abstract}

It is generally believed that a Wigner Crystal in single layer
graphene can not form because the magnitudes of the Coulomb
interaction and the kinetic energy scale similarly with decreasing
electron density. However, this scaling argument does not hold for
the low energy states in bilayer graphene. We consider the
formation of a Wigner Crystal in weakly doped bilayer graphene
with an energy gap opened by a perpendicular electric field. We
argue that in this system the formation of the Wigner Crystal is
not only possible, but different phases of the crystal with very
peculiar properties may exist here depending on the system
parameters.

\end{abstract}
 \pacs{73.20.Qt, 73.22.Pr, 81.05.ue}
\maketitle

{\it 1. Introduction. --} The observation of a Wigner
crystal~\cite{Wigner34,Bonsall79,Tanatar89,Drumond09}, a
solidified phase of a conducting electron Fermi liquid, is a
challenging task even in conventional
metals~\cite{Helium79,EAndrey88,Yoon99}. Finding, whether this
elusive electron solid may exist in novel materials, like
graphene, constitutes an additional challenge. Simple scaling
analysis shows that Wigner crystallization in single layer
graphene is unlikely~\cite{Balatsky06}. Much richer is the issue
of crystallization in bilayer graphene. If the gap in the bilayer
electron spectrum is not opened, screening of the Coulomb
interaction~\cite{Hwang08} prevents the formation of the crystal,
as we discuss below~\cite{ endnote}. However, and this is the main
result of this paper, if the gap is opened due to an interlayer
voltage, the crystallization is not only possible, but depending
on the density of electrons in the conduction band there should
exist two very distinct phases of the crystal.

Interaction induced phases of undoped bilayer graphene have been
addressed by numerous publications~\cite{Nilsson06, Min08,
McDonald10, Vafek10, NandkishorePRL10, NandkishorePRB10,
Lemonik10, Throckmorton12, Throckmorton14}. The considered effects
include the spontaneous ferromagnetic and/or pseudospin
polarization~\cite{Nilsson06, Min08} or the spontaneous gap
opening in the electron's spectrum~\cite{NandkishorePRL10,
NandkishorePRB10}, but not the spontaneous translation symmetry
breaking. The spatial in-plane charge inhomogeneity is hard to
expect in undoped bilayer graphene. In this paper we consider the
breaking of translation symmetry in the form of an electron
crystal in the case of a weakly doped conduction band in gapped
bilayer graphene. A gate tunable doping level is obtained
routinely in both single- and bilayer
graphene~\cite{Novosel05,Zhang05}.

Generally, Wigner crystallization takes place when the lowering of
electrons' repulsion energy in the crystal phase wins over the
rise of the kinetic energy caused by the restricted motion on the
lattice~\cite{Wigner34}. Both the kinetic energy and the screened
interaction behave highly nontrivially in bilayer graphene with
the interlayer voltage induced gap. After the gap is opened,
graphene becomes an insulator and electrons' repulsion at very
large distances becomes the usual Coulomb-law. However, at least
one of the Wigner crystal phases, which we consider, exists for
the inter-electron distances where the electrons' interaction is
well approximated by a logarithmic repulsion, reminiscent of the
vortex interaction in type II
superconductors~\cite{Abrikosov57,SBook}.

\begin{figure}
\includegraphics[width=7.cm]{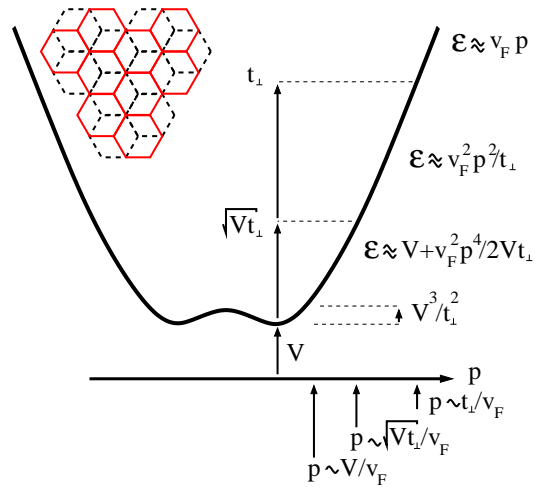}
\caption{Schematic drawing of the different regimes for the
kinetic energy dispersion $\varepsilon(p)$ in bilayer graphene
with a gap opened by an interlayer voltage $V$ and with interlayer
hopping energy $t_\perp$. Momentum is calculated from either $K$
or $K'$ point. The inset shows the bilayer graphene lattice with
Bernal AB-stacking. } \label{fig:1}
\end{figure}

Deep in the stable crystal phase, which is the only regime
accessible analytically, the dominant repulsion of electrons
favors the triangular lattice~\cite{Bonsall79} with only small
fluctuations around it. The kinetic energy, which has a unique
form for electrons in bilayer graphene, is responsible in this
regime for quantum fluctuations and small distortions of the
classical lattice. These fluctuations may or may not respect the
symmetries of the original triangular lattice. For example, the
lattice symmetry is preserved at the quantum level in a Wigner
crystal with quadratic dispersion ($\eps\sim p^2/2m$) but is
broken in 2-dimensional semiconductors with strong spin-orbit
interaction~\cite{Berg12,Silvestrov14}. As we will show, both of
these possibilities are realized in gapped bilayer graphene at
different electron densities.

Applying an interlayer voltage, besides opening the gap, leads to
a peculiar single-electron dispersion with several regions of
different scaling behavior as a function of momentum (see
Fig.~\ref{fig:1}). Consequently, the two phases of the crystal,
which we predict in this work, are distinguished by different
kinetic energy dispersions at different electron densities. While
the doping level in the conduction band is lowered,
the dilute electron gas
crystalizes into what we call an intermediate density crystal
phase with a quartic electron dispersion, $\eps (p)\propto p^4$.
This anharmonic kinetic energy makes it difficult to describe the
quantum fluctuations of the crystal. We use the self-consistent
mean-field approximation to calculate the effective phonon modes
in this case, which may be reasonable even in the absence of a
small parameter. The quantum corrections in the case of $\eps
(p)\propto p^4$ obviously preserve the symmetries of the
triangular crystal lattice.

\begin{figure}
\includegraphics[width=8.cm]{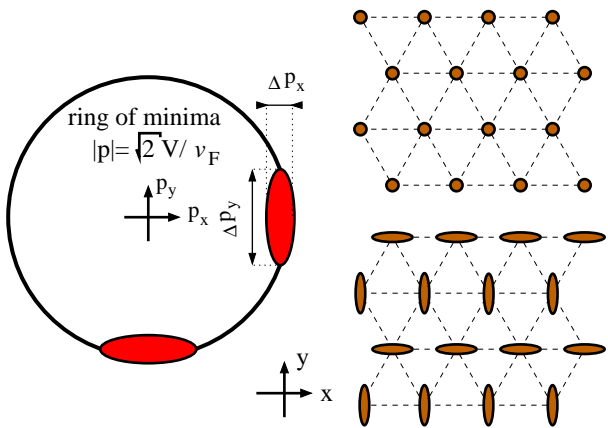}
\caption{Visualization of the  electronic density in a Wigner
crystal with Mexican-hat shaped kinetic energy dispersion in the
momentum(left) and coordinate(bottom-right) representation
according to Ref.~\cite{Silvestrov14}. Top-right - the Wigner
crystal at the intermediate electron density, where fluctuations
preserve the triangular lattice symmetries.} \label{fig:2}
\end{figure}

When reducing the density of electrons further their energies get
close to the bottom of the sombrero-like spectrum characteristic
of graphene with an interlayer voltage~\cite{McCann06}. This
dispersion relation is reminiscent of the one for electrons with
Rashba spin-orbit interaction~\cite{Rashba60}. The Wigner
crystallization in a two-dimensional electron gas with strong
spin-orbit interaction was investigated in
Refs.~\cite{Berg12,Silvestrov14}. The predictions of
Ref.~\cite{Silvestrov14}, where the long-range interaction between
electrons was assumed, may be applied to bilayer graphene almost
without modifications. The main effect for the fluctuations in
this low-density regime for bilayer graphene is an asymmetric
(cigar-shape) density profile in real as well as momentum space
which breaks the symmetries of the original triangular lattice.
The fact that the two low energy phases of the electron crystal
which we find have different symmetries rules out the possibility
of a smooth crossover between them (see also Fig.~(\ref{fig:2})).

Taking into account properly the screening of electron-electron
interaction is crucial for the correct description of Wigner
crystallization in bilayer graphene. We describe below different
screening regimes for the case of a parametrically small gap in
bilayer graphene and give more details on the screening in the
Appendix.

{\it 2. Bilayer graphene.--} The Bloch Hamiltonian for the Bernal
stacked bilayer graphene  in the vicinity of the $K$ point is
given by the matrix~\cite{McCann06} (we neglect the hopping
elements leading to the small trigonal warping terms)
 \begin{align}\label{HamK}
{\cal H}_{\vec p}=\left(\begin{array}{cc}
-V \ \ \ \ \ \ v_F {\tilde p} \ \ \ \ 0 \ \ \ \ \ 0\\
 v_F {\tilde p}^*  \ \ -V \ \ \ \ t_\perp \ \ \ \ \ 0\\
\ \ \ \ 0 \ \ \ \ \ \ \ \ t_\perp \ \ \ \ \ V \ \ \ \ v_F {\tilde p}\\
\ \ 0 \ \ \ \ \ \ \ \ 0 \ \ \ \ v_F {\tilde p}^* \ \ \ V
\end{array}\right)\ .
 \end{align}
Here $\vec p$ is the momentum calculated from the $K$ point and
${\tilde p}=p_x+ip_y$. The hopping matrix element between two
vertically aligned carbon atoms $t_\perp\approx 0.39\, {\rm eV}$
is small compared to the interlayer matrix element $t\approx 2.8\,
{\rm eV}$, the latter entering Eq.~(\ref{HamK}) through the Fermi
velocity $\hbar v_F=3dt/2$ with $d\approx 1.42$\AA \ being the
distance between two nearest in-plane carbon atoms. The layer
potential $\pm V$ is regulated by the external gates. Each single
electron state is doubly degenerate due to spin and electrons with
momenta close to the $K'$ point are described by the Hamiltonian
${\cal H}^*_{-\vec p}$.

Diagonalization of ${\cal H}_{\vec p}$ gives the particle-hole
symmetric spectrum
 \bq\label{eVexact}
\eps^2 = V^2 +v_F^2 p^2 +\fr{t_\perp^2}{2} \pm \sqrt{ v_F^2
p^2 (4V^2+ t_\perp^2) +\fr{t_\perp^4}{4}} .
 \ee
For an interlayer voltage small compared to the interlayer hopping
matrix element $V/t_\perp\ll 1$, the two crystal phases exist at
parametrically different electron densities. The ratio
$V/t_\perp$, which is a small parameter in our estimates, may be
tuned experimentally. Assuming $V\ll t_\perp\ll t$, we find
several distinct regimes of the spectrum Eq.~(\ref{eVexact}),
 \begin{align}\label{eVapprox}
{\rm I.}& \ \ \eps\approx v_F p \ \ \
{\rm for}  \ \ \ v_F p\gg t_\perp \ ,\\
{\rm II.}& \ \ \eps\approx {v_F^2 p^2}/{t_\perp}  \ \ \
{\rm for}  \ \ \ \sqrt{V t_\perp}\ll v_F p\ll t_\perp \ , \nonumber\\
{\rm III.}& \ \ \eps\approx V+ \fr{v_F^4 p^4}{2V t_\perp^2} \ \ \
{\rm for}  \ \ \ V\ll v_F p\ll \sqrt{V t_\perp} \ , \nonumber
\\
{\rm IV.}& \ \ \eps\approx V -2V\fr{v_F^2
p^2}{t_\perp^2}+\fr{v_F^4 p^4}{2V t_\perp^2} \ \ \ {\rm for} \ \ \
v_F p\sim V \ . \nonumber
 \end{align}
We show here only the low energy positive branch of solutions
Eq.~(\ref{eVexact}). Replacing the momentum by the inverse typical
distance between electrons, $p \sim\sqrt{n}$, one finds the
electron density $n$ assigned to each energy regime I-IV.

As we discuss below, Wigner crystallization is possible only in
the lowest energy/density regimes III and IV of
Eq.~(\ref{eVapprox}).

{\it 3. Screening of the Coulomb interaction.--} The different
energy dispersion regimes in Eq.~(\ref{eVapprox}) lead to a
different ability of bilayer graphene to screen the coulombic
electron-electron interaction at different length scales. We give
a detailed description of the screening in the Appendix, and
present here only the results.

First, at the highest energies or electron densities, the two
graphene sheets are approximately decoupled, leading to a
single-layer-like Coulomb interaction~\cite{KotovRMP12},
 \bq\label{screening1}
U(r) ={e^2}/{\epsilon \, r},
 \ee
where the polarization of both the substrate and the graphene
flake contribute similarly to the dielectric constant
$\epsilon\sim 1$ (see Appendix).

The long-range Coulomb interaction in undoped intrinsic bilayer
graphene is fully screened in the second regime II of
Eq.~(\ref{eVapprox}) with quadratic dispersion and negligible gap,
$\eps\propto p^2$. The random-phase-approximation calculation of
Ref.~\cite{Hwang08} here gives
 \bq\label{screening2}
U(r\gg 1/q_{\rm TF})\sim {e^2}/{(q_{\rm TF}r)^2 r} \ ,
 \ee
where the Thomas-Fermy screening wave-vector $q_{\rm TF}\sim
t_\perp/\hbar v_{\rm F}$ (see Appendix for details). Note that in
two dimensions screening of the charge is not exponential but a
power law, leading to a  $\sim 1/r^3$ interaction, as it is in
Eq.~(\ref{screening2}).


At distances $r> \hbar v_{\rm F}/\sqrt{V t_\perp}=2dt/3 \sqrt{V
t_\perp}$ (regime III) bilayer graphene behaves like a
two-dimensional insulator due to the gap in the spectrum, with a
large and momentum dependent dielectric constant, leading to (see
Appendix)
 \bq\label{screening3}
U(r\gg \hbar v_{\rm F}/\sqrt{V t_\perp}) = (3V/4)\ln(\hbar v_F/Vr) \ .
 \ee
Using this potential is enough for a quantitative description of
the Wigner crystallization and of the properties of the crystal in
the region III of Eq.~(\ref{eVapprox}). It is also sufficient for
the qualitative description of the transition between the two
crystal phases at $r\sim \hbar v_F/V$. Only at much larger
inter-electron distances (deep inside the regime IV) the effect of the
graphene polarization become negligible and the interaction takes
the form
 \bq\label{screening4}
U(r\gg \hbar v_{\rm F}/V) = {e^2}/{\epsilon_0 r } \ ,
 \ee
with $\epsilon_0$ being the substrate dielectric constant (see
Appendix).

Understanding the screening behavior
Eqs.~(\ref{screening1}-\ref{screening4}) of the electron-electron
interaction is crucial for understanding the possibility of Wigner
crystallization in bilayer graphene. For two-dimensional electron
gases formed in usual semiconductor heterostructures, the
interaction between electrons is of the Coulomb form, $U(r)\sim
1/r$, and the kinetic energy is quadratic in momentum, $\eps\sim
p^2\sim \hbar^2/\Delta r^2$. Here $\Delta r$ is the electrons'
quantum mechanical position uncertainty, which at the melting
transition is of the same order as the typical distance between
electrons. With lowering the electron density the kinetic energy
decays faster than the typical electron interaction thus making
the crystalline phase energetically favorable~\cite{Wigner34,
Bonsall79, Tanatar89, Drumond09}. On the contrary, in the single
layer graphene the electron energy $\eps=v_{\rm F}p\sim \hbar
v_{\rm F}/\Delta r$ scales at low electron densities similarly as
the Coulomb interaction energy, making the Wigner crystallization
unlikely~\cite{Balatsky06}.

In ungapped bilayer graphene the electron dispersion relation,
Eq.~(\ref{eVapprox}) II, becomes quadratic in momentum like for
usual semiconductors. However, the Wigner crystal can not exist
here because of the strong screening from the filled valence band
leading to a $U(r)\sim 1/r^3$ interaction, Eq.~(\ref{screening2}).
The authors of Ref.~\cite{Balatsky2010} have considered the
possibility of a CDW (charge-density wave) instability in doped
ungapped bilayer graphene. However, the screening of long-range
interaction (see Eq.~(\ref{screening2}) and Ref.~\cite{Hwang08})
was not taken into account in Ref.~\cite{Balatsky2010} and
therefore their results are not applicable to the low density
electron phase.

Only in gapped bilayer graphene, where the kinetic energy is
sufficiently suppressed, Eq.~(\ref{eVapprox}) III and IV, and the
interaction is strong enough,
Eqs.~(\ref{screening3},\ref{screening4}), crystallization of a
dilute electron gas becomes possible.

{\it 4. Existence of the Wigner crystal in gapped bilayer.--} The
electron crystal in bilayer graphene at the densities where
crystallization is possible is thus described by a Hamiltonian
 \bq\label{HamWigner}
{\cal H }= \sum_i {\cal{H}}_0(p_i) +\sum_{i<j} U(|\vec R_{ij}
+\vec r_{ij}|)
 \ ,
 \ee
where the single-electron Hamiltonian is determined by its
eigenvalues ${\cal H}_0(p)$, (cf. Eq.~(\ref{eVapprox})) and the
potential $U(r)$ (Eqs.~(\ref{screening1} - \ref{screening4})) in
the region of our interest is best approximated by the logarithmic
formula Eq.~(\ref{screening3}). In the crystal phase the
electrons' displacements $\vec r_i$ from their equilibrium
positions $\vec R_i$ should be small compared to the lattice
constant, which we denote by $a$. Also $\vec R_{ij}=\vec R_i-\vec
R_j$ and $\vec r_{ij}=\vec r_i-\vec r_j$.

Consider first the higher density Wigner crystal phase with single
electron energies of the form Eq.~(\ref{eVapprox})~III. For small
displacements around the equilibrium electron positions,
Hamiltonian Eq.~(\ref{HamWigner}) now takes the form
 \bq\label{HamIII}
{\cal H}_{\rm III}=\lambda\sum_j \vec p_j^4
+\sum_{i<j,\alpha,\beta} r_{ij}^\alpha r_{ij}^\beta
u^{\alpha\beta}_{ij} \ ,
 \ee
where $\lambda =v_F^4/(2V t_\perp^2)$, $r_{ij}^\alpha$ is the
$\alpha$-component of vector $\vec r_{ij}$ and components of the
tensor $u^{\alpha\beta}_{ij}\sim V/R_{ij}^2$ are found via the
small displacement expansion of the potential
Eq.~(\ref{screening3}). Electrons fluctuate around their
equilibrium positions with some typical amplitude $\Delta r$. To
ensure the crystal stability, two obvious conditions should be
met. First, the amplitude of quantum fluctuations should be small
compared to the lattice spacing, $\Delta r \ll a$. Second, the
crystallization reduces the interaction energy, but raises the
fluctuation energy of confined electrons. The crystal phase is
stable if this decrease of interaction energy exceeds the gain in
the fluctuation one.

The two terms in Eq.~(\ref{HamIII}) give comparable contributions
to the ground state quantum fluctuation energy. This allows us to
find the typical displacement $\Delta r^2$ from
 \bq\label{displacement}
\lambda {\hbar^4}/{\Delta r^4}\sim ({V}/{a^2})\Delta r^2 \ .
 \ee
Requiring the smallness of either the amplitude of fluctuations or
the fluctuation energy now gives
 \bq\label{liquid-crystal}
a\gg {dt}/\sqrt{V t_\perp} \ .
 \ee
This determines an upper bound for the electron density, $n\sim
1/a^2$, in a stable crystal. Electrons in the regime III Eq.
(\ref{eVapprox}) resemble Abrikosov vortices in type~II
superconductors~\cite{Abrikosov57}, which repel each other
logarithmically and are known to crystallize into the triangular
lattice~\cite{SBook}.

For $V\ll t_\perp$ electron gas crystallization
Eq.~(\ref{liquid-crystal}) takes place at higher electron's
densities than needed for the Fermi liquid symmetry-breaking
transition suggested in Refs.~\cite{BergPRB15,MacdonaldPRB14}.

With further increasing the distance between electrons one needs
to take into account the (negative)quadratic term in the energy
dispersion Eq.~(\ref{eVapprox})~IV, which happens at $\Delta r\sim
d\, t /V$. Since at the transition between two crystal phases both
terms $\sim p^2$ and $\sim p^4$ are of the same order of
magnitude, we still can use here Eq.~(\ref{displacement}) to find
the relation between $\Delta r$ and $a$. Thus we find at the
transition
 \bq\label{crystal-crystal}
a\sim dt \, t_\perp/V^2 \ .
 \ee
This corresponds to a $t_\perp^3/V^3\gg 1$ times lower density, as
needed for the liquid-to-crystal transition
Eq.~(\ref{liquid-crystal}). We will return to the discussion of
the Wigner crystal phase for the case of a Mexican-hat electron
spectrum later.

Finding the accurate positions of the phase transitions
characterized by the electron densities described by
Eqs.~(\ref{liquid-crystal}) and (\ref{crystal-crystal}) may be
done only numerically. However, our estimates are enough to prove
the existence of such liquid-to-crystal~(\ref{liquid-crystal}) and
crystal-to-crystal~(\ref{crystal-crystal}) phase transitions in
bilayer graphene for a sufficiently weak interlayer voltage, $V\ll
t_\perp$.

{\it 5. Mean field approach to quantum fluctuations.--} The
displacement Hamiltonian Eq.~(\ref{HamIII}) with the $\lambda p^4$
single particle energy, does not support even small amplitude
harmonic vibrations, which
would lead to the phonon modes. One way to treat this Hamiltonian
approximately is to perform the mean field decomposition
 \bq\label{meanfield}
\lambda\sum_j \vec p_i^4 \rightarrow \lambda\sum_j (2\vec p_i^2
\langle\vec p^2\rangle+ 4p_i^\alpha p_i^\beta\langle p^\alpha
p^\beta\rangle) \ ,
 \ee
where the onsite expectation values $\langle\vec p^2\rangle$ and
$\langle p^\alpha p^\beta\rangle$ should be found
selfconsistently.

Using Eq.~(\ref{meanfield}) makes the displacement Hamiltonian
Eq.~(\ref{HamIII}) exactly solvable. The true $\lambda p^4$
kinetic energy may then be taken into account perturbatively. The
advantage of choosing Eq.~(\ref{meanfield}) as a zeroth order
approximation is that it leads to the perturbation theory
expansion with vanishing first order diagrams in the phonon
interaction. However, the resulting series, although starting from
the second order, has no obvious small parameter.

Due to the triangular symmetry of the crystal we have $\langle
p^\alpha p^\beta\rangle =\delta_{\alpha\beta} \langle\vec
p^2\rangle/2$. Thus using Eq.~(\ref{meanfield}) together with
Hamiltonian Eq.~(\ref{HamIII}) is equivalent to introducing a
usual quadratic dispersion with the effective mass
 \bq\label{meff1}
{1}/{(2 m_{\rm eff})}= 4\lambda \langle\vec p^2\rangle \ .
 \ee
In order to find the average value $\langle\vec p^2\rangle$, we
may use the virial theorem, which states that the energy of a
system of harmonic oscillators ($\hbar \omega_k/2$ per mode) is
split equally between the kinetic and potential energy terms,
$\fr{1}{2 m_{\rm eff}}{\langle\vec p^2\rangle} =
\fr{\hbar}{4}\sum_\vec k \fr{\omega_k}{N} $. The phonon frequency
$\omega_k$ with wave vector $\vec k$ depends itself on $m_{\rm
eff}$. The mass independent combination is $m_{\rm
eff}\omega_k^2\sim V/a^2$. Therefore it is convenient to rewrite
Eq.~(\ref{meff1}) as
 \bq\label{meff2}
\fr{1}{m_{\rm eff}^{3/2}}= 4\lambda\hbar\sqrt{m_{\rm
eff}}\sum_\vec k \fr{\omega_k}{N} =\alpha \fr{\lambda\hbar V}{a^2}
\ .
 \ee
The coefficient $\alpha\sim 1$ here may be found numerically.

Approximations Eqs.~(\ref{meanfield})-(\ref{meff2}) should give a
reasonably good description of the typical displacement eigenmodes
of the Hamiltonian Eq.~(\ref{HamIII}). However, it is unclear, how
the strong interaction between phonons would modify for example
the spectrum of low energy excitations.

{\it 6. Mexican-hat potential.--} Existing previous investigations
have been concentrated on the Wigner crystallization starting from
the Fermi liquid
phase~\cite{Wigner34,Bonsall79,Tanatar89,Drumond09}.
Interestingly, in bilayer graphene in addition to the
liquid-to-crystal transition we predict a second low density
transition between two solidified phases, indicating the lattice
reconstruction caused by the Mexican-hat shaped kinetic
energy~Eq.~(\ref{eVapprox})~IV.

Freezing of the Fermi liquid may be seen in transport
experiments~\cite{Yoon99}. To detect the second transition one may
search for the change of the symmetry of the lattice (seen e.g. in
the photon reflection). However, probably the easiest way, for
which graphene is almost ideally designed (see e.g. the
experiments~\cite{EisensteinPRB2010,YoungPRB2012}), will be to
measure the singularity in the doping dependance of the
differential capacitance at the transition. Doping of (exfoliated)
graphene is achieved by applying a voltage between the conducting
substrate and the graphene flake separated by an insulating layer.
At a given electron density the value of the voltage depends on
the interaction energy per electron in the Wigner crystal.
Consequently, the differential capacitance carries the information
about the phase of the Wigner crystal and about the transition
between phases.

As we mentioned already, the properties of the lowest density
Wigner crystal phase in bilayer graphene are similar to that of
the Wigner crystal in a two-dimensional electron gas with strong
Rashba spin-orbit interaction, investigated recently by one of the
authors~\cite{Silvestrov14}. The electron density (coordinate and
momentum representation) found in Ref.~\cite{Silvestrov14} is
shown in Fig.~\ref{fig:2}. Electrons' crystallization into the
intermediate density phase (dispersion $\eps\propto p^4$) of the
Wigner crystal breaks the continuous symmetry of the Fermi liquid
to the discrete symmetry ($D6$) of the triangular lattice. We
expect quantum fluctuations in this phase to preserve the lattice
discrete symmetries. On the contrary, the triangular lattice
symmetry is broken by the fluctuations in the lowest density
phase, where one has to take into account the multiple minima of
the Mexican-hat shaped energy dispersion. These different
symmetries of the lattice at different densities prove the
existence of the quantum phase transition.

The regime Eq.~(\ref{eVapprox})~IV exhibits a degenerate minimum
at the ring of momenta $|\vec p|=\sqrt{2}V/v_F$. Since the
uncertainty principle couples coordinates and momenta, broken
spatial rotational symmetry in the crystalline phase makes
different directions in the momentum space also inequivalent. Each
crystal electron now picks up its own position (different for
different electrons) at the ring of minima. For any choice of the
set of minima, vibrations normal to the ring $|\vec
p|=\sqrt{2}V/v_F$ give parametrically the largest contribution to
the fluctuation energy. Minimization of the zero point energy due
to these fluctuations for the ensemble of individual electron
positions on the ring results in the crystal shown in
Fig.~\ref{fig:2}~\cite{Silvestrov14}. The low energy excitation
modes of the Wigner crystal with a Mexican hat shaped kinetic
energy, associated e.g. with the vibrations along the ring of
minima, $|\vec p|=\sqrt{2}V/v_F$, or with the valley and spin
flips, can not lead to a substantial change of the electron
density distribution.

{\it 7. Conclusions.--} Our main result in this paper is the
prediction of the existence of Wigner crystalline phases in
lightly doped bilayer graphene subject to an interlayer voltage.
This is in contrast to single-layer graphene and bilayer graphene
without a gap, where scaling arguments (together with screening
properties for bilayer graphene) prove the absence of the
crystallization~\cite{Balatsky06}.

Moreover, we predict the existence of two distinct crystal phases
at different electron densities, having different symmetries and are
separated by a quantum phase transition. We suggest differential
capacitance measurements for the experimental verification of the
transition.

Further investigations of Wigner crystals with non-quadratic
kinetic energy may be done numerically via the Quantum Monte Carlo
method. At least for the case of the quartic energy dispersion,
$\eps\propto p^4$, this may be not more difficult than the
standard calculation~\cite{Tanatar89,Drumond09}.

The two phases of the Wigner crystal predicted in this paper exist
if the voltage between the graphene layers is small compared to
the interlayer coupling, $V\ll t_\perp$. Assuming $V=t_\perp/10$
we estimate from
Eqs.~(\ref{liquid-crystal},~\ref{crystal-crystal}) the
liquid-to-solid ($ls$) and the solid-to-solid ($ss$) transitions
to appear at $a_{ls}\approx 23\, d$ and $a_{ss}\approx 720\, d$,
or at the electron densities $n_{ls}\approx 1.1\cdot 10^{13}{\rm
cm}^{-2}$ and $n_{ss}\approx 1.1\cdot 10^{10}{\rm cm}^{-2}$. We
must mention however, that in the usual electron gas with
parabolic dispersion consideration similar to ours overestimates
the transition density by a large pure numerical
factor~\cite{Tanatar89,Drumond09}.

Concerning other effects potentially affecting the
crystallization: Disorder may be ignored, since electron's mean
free path in bilayer graphene may be as large as
micrometers~\cite{Schoeneberger2015}. Temperature, to play a role,
should be comparable to (e.g. kinetic) electron's energy. For the
low density solid-to-solid transition this gives $T\sim
V^3/t_\perp^2\approx 5{\rm K}$ (for $V=t_\perp/10$). As
usual~\cite{EAndrey88}, a quantizing magnetic field will simplify
the observation of a (Skyrme-)Wigner~\cite{WignerMagnetic,
SkyrmeMagnetic} crystal in bilayer graphene. However, the magnetic
field will also destroy the low density crystal phase from
Fig.~\ref{fig:2}.

\begin{acknowledgments}

{\it Acknowledgements.--} Discussions with A.~V. Balatsky, E.
Bergholtz, O. Entin-Wohlman, S.~Park, I.~V.~Protopopov,
R.~Ramazashvili and O.~P. Sushkov are greatly acknowledged. This
work was supported by the DFG grant RE~2978/1-1.

\end{acknowledgments}



\vspace{1.cm}

\section{Appendix: Screening in gapped bilayer graphene}

Here, we consider the static screening of the Coulomb interaction
between electrons in intrinsic(undoped) bilayer graphene with an
interlayer voltage. A detailed discussion of the interaction
effects in graphene may be found e.g. in the
review~\cite{KotovRMP12}. However we are not aware of any
publication emphasizing the absence of interaction corrections in
the dielectric constant of gapped bilayer graphene and especially
the existence of the intermediate regime for the interaction
Eq.~(\ref{Urlog}) for $V\ll t_\perp$.

The standard approach to screening of the electrostatic potential
is by introducing the momentum dependent dielectric constant
$\epsilon(q)$ via
 \bq\label{epsilon}
U(q)=\fr{2\pi e^2}{\epsilon_0 q}\ \rightarrow \ \fr{2\pi
e^2}{\epsilon(q) q} \ .
 \ee
Here, the bare dielectric constant $\epsilon_0$ for the case of
exfoliated graphene on silicon-oxide is
$\epsilon_0=(\epsilon_{{\rm SiO_2}}+1)/2$~\cite{SilvestrovPRB08}
and $\epsilon_{SiO_2}\approx 4.2$. In suspended graphene obviously
$\epsilon_0=1$. The dielectric function is usually calculated in
the random phase approximation (RPA), yielding
 \bq\label{epsilonPi}
\epsilon(q)=\epsilon_0[1-U(q)\Pi(q)] \ ,
 \ee
where the polarization function $\Pi(q)$ is found in a
single bubble approximation.

In order to have a finite dielectric constant at low momenta the
polarization function should vanish at small $q$ as
$\Pi(q\rightarrow 0)\sim q$. This indeed happens in a single layer
graphene, where the static dielectric constant in the RPA
approximation takes the form
 \bq\label{VRPA}
U_{\rm RPA} = \fr{1}{\epsilon_{\rm RPA}}\fr{e^2}{r} \ , \
\epsilon_{\rm RPA} =\fr{\epsilon_{{\rm SiO_2}}+1}{2}
+\fr{\pi}{2}\fr{e^2}{\hbar v_F}\ .
 \ee
For the case of bilayer graphene in the regime I of Eq.~(\ref{eVapprox})
of the main text one should simply double the interaction ($\sim
e^2$) term in $\epsilon_{\rm RPA}$ (\ref{VRPA}) and use this
formula for $r< \hbar\, v_F/t_\perp$.

The result Eq.~(\ref{VRPA}) is already surprising. Graphene is
only a two-dimensional sheet of atoms. How can it show the same
effect on screening of a long-range interaction potential as a
three-dimensional bulk of ${\rm SiO_2}$? This may happen only
because graphene has no bandgap and consequently a much higher
polarizability.

The situation is even more interesting in bilayer graphene. In the
ungapped case the spectrum of the bilayer consists of two
parabolic bands touching each other at the $K$ ($K'$) point. This
means that the density of states around the Fermi energy in
intrinsic bilayer graphene is constant
 \bq
\left. \fr{dn}{d\eps}\right|_{\eps\approx 0}=const \neq 0 \ ,
 \ee
similar to that in a usual two-dimensional metal. This results in
an even larger polarizability than for single layer graphene,
sufficient to develop a full screening of the electric charge, as
was shown in the one-loop calculation in Ref.~\cite{Hwang08}. In
this case the polarization operator $\Pi(q)$ turns out to be a
constant independent of $q$ and the interaction potential in the
momentum representation Eq.~(\ref{epsilon}) takes the form
 \bq\label{screeningTF}
U(q) = \fr{2\pi e^2}{\epsilon_0 q}\rightarrow \fr{2\pi
e^2}{\epsilon_0 (q+q_{\rm TF})} \ ,
 \ee
where the Thomas-Fermi screening wave vector is~\cite{Hwang08}
 \bq\label{qTF}
q_{\rm TF}=\fr{2 t_\perp e^2}{\epsilon_0 \hbar^2 v_F^2}\ln 4 \ .
 \ee
The charge of the electron in Eq.~(\ref{screeningTF}) is fully
compensated by the cloud of charges induced in graphene at a
distance $\sim 1/q_{\rm TF}$. Since this is a two-dimensional
charge cloud, the potential is not fully screened, but rather
decays (in plane) as a power law $U(r\gg 1/q_{\rm TF})\sim 1/r^3$.

In bilayer graphene with an interlayer voltage $V$, described by
the Hamiltonian Eq.~(\ref{HamK}), the result
Eq.~(\ref{screeningTF}) is valid as long as one may neglect the
gap $\sim V$ in the electron's spectrum, i.e. at $q\gg \sqrt{V
t_\perp}/\hbar v_F$. For smaller momenta (larger distances between
electrons) one should consider the graphene sheet as a narrow-gap
two-dimensional insulator. The polarization function $\Pi(q)$ here
decreases with decreasing $q$ as $\Pi(q)\sim q^2$ (see the
calculation below). Eventually at very small $q$ the polarization
function contribution to the dielectric constant becomes
negligible, i.e. $\epsilon(q)=\epsilon_0$ in Eq.~(\ref{epsilon}).
This means that at largest distances the screening of the
electron's interaction is fully determined by the bulk
three-dimensional dielectric below and above the graphene flake,
$U(r\rightarrow\infty) = e^2/\epsilon_0 r$. The transition from
the fully screened, $U\sim 1/r^3$, to the unscreened Coulomb
interaction does not happen instantaneously, but rather proceeds
continuously in the parametrically wide region of inter-electron
distances, $\hbar v_F/\sqrt{t_\perp V}\ll r\ll \hbar v_F/V$. Below
we describe the behavior of the screened potential $U(r)$ at these
intermediate distances.

To describe quantitatively the evolution of the electron's
interaction from the fully screened to the unscreened regime, we
consider the calculation of the polarization function $\Pi(q)$.
First we notice that the Thomas-Fermi screening Eq.~(\ref{qTF})
and transition between quadratic (II) and linear (I) spectrum
regimes in Eq.~(\ref{eVapprox}) of the main text take place at the
same momentum $\sim q_{\rm TF}$, since in graphene $e^2/(\hbar
v_F)\sim 1$. This means that we may use a simplified two-band
Hamiltonian for a reliable description of the polarization,
instead of the full four-band Hamiltonian Eq.~(\ref{HamK}),
cf.~\cite{Hwang08},
 \begin{align}\label{Heff}
{\cal H}_{eff}=\left(\begin{array}{cc}
\ \ \ \ \ \ -V \ \ \ \ \ -{\tilde p}^2 v_F^2/t_\perp \\
 -{\tilde p}^{*2} v_F^2/t_\perp  \ \ \ \ V
\end{array}\right)\ ,
 \end{align}
where ${\tilde p}=p_x+ip_y$. The two eigenvalues of the two-band
Hamiltonian are
 \bq\label{eps2band}
\eps_{p\pm} =\pm \eps_p \ , \ \eps_p=\sqrt{V^2 +
p^4v_F^4/t_\perp^2 } \ ,
 \ee
which reproduce correctly the spectrum of the four-band
Hamiltonian Eq.~(\ref{HamK}) in the regimes II and III of
Eq.~(\ref{eVapprox}). The lowest energy Mexican-hat spectrum,
Eq.~(\ref{eVapprox}) regime  IV, may be found only from the
four-band model. However, as we will see, the two-band
approximation Eq.~(\ref{Heff}) leads to the correct form of the
polarization function $\Pi(q)$ even for the momentum transfer $q$
corresponding to the lowest energy regime IV in
Eq.~(\ref{eVapprox}). The two positive- and negative-energy
eigenfunctions of ${\cal H}_{eff}$ in Eq.~(\ref{Heff}) are
 \bq\label{psi+}
\psi_+ =\fr{
1}{\sqrt{{ p}^4
v_F^4/t_\perp^2+(V+\eps_p)^2}}\left(\begin{array}{cc}
{\tilde p}^2 v_F^2/t_\perp \\
 -(V+\eps_p)
\end{array}\right) \ ,
 \ee
and
 \bq\label{psi-}
\psi_- =\fr{ 1}{\sqrt{{ p}^4
v_F^4/t_\perp^2+(V+\eps_p)^2}}\left(\begin{array}{cc}
V+\eps_p\\
{\tilde p}^{*2} v_F^2/t_\perp
\end{array}\right) \ .
 \ee

The zero temperature polarization function in the single bubble
approximation may now be written as
 \bq\label{Pi(q)}
\Pi(q) =-g\int \fr{d^2 p}{(2\pi\hbar)^2}\left[ \fr{|\psi_+^\dagger
\psi_-'|^2}{\eps_{\vec p+}-\eps_{\vec p'-}} + \fr{|\psi_+^\dagger
\psi_-'|^2}{\eps_{\vec p'+}-\eps_{\vec p-}}\right] \ .
 \ee
Here $\hbar\vec q =\vec p -\vec p'$, $\psi'=\psi(p')$ and the
degeneracy factor $g=4$ accounts for two valleys and two spin
orientations. For the gapless case, $V=0$, formulas for $\psi_+$
and $\psi_-$ are greatly simplified and Eq.~(\ref{Pi(q)})
coincides with Eq.~(4) of Ref.~\cite{Hwang08}.

Moreover, in the case of vanishing $V$ the overlap of two
eigenvectors depends only on the ratio $p/q$ and the angle between
two momenta, $|\psi^\dagger_{ +}\psi'_{ -}|^2 = 1- (\vec p \vec
p')^2/p^2p'^2$, while the energy in this case is simply quadratic
in momentum $\eps_{\vec p\pm}\propto \pm p^2$. As a result the
polarization function Eq.~(\ref{Pi(q)}) in the limit $V=0$ turns
out to be a constant independent of the momentum transfer
$q$~\cite{Hwang08}
 \bq\label{Pi_V=0}
\Pi(q) \approx \Pi_{V=0} = -\fr{\ln 4}{\pi} \fr{t_\perp}{\hbar^2
v_F^2} \ ,
 \ee
leading to the Thomas-Fermi screening Eqs.~(\ref{screeningTF},
\ref{qTF}).

The situation is different in the case of a finite interlayer
voltage, $V\neq 0$, and a very small momentum transfer, $ q\ll
\sqrt{t_\perp V}/\hbar v_F$. The integral over momentum in this
limit comes from the region $p\sim \sqrt{t_\perp V}/v_F$. This
means that typically $p\gg q$ and the overlap of two different
eigenfunctions of almost the same Hamiltonians, ${\cal
H}_{eff}(p)$ and ${\cal H}_{eff}(p')$ in Eq.~(\ref{Heff}),
vanishes as $|\psi^\dagger_{ +}\psi'_{ -}|^2\propto q^2$. This
leads to the quadratic in small momentum transfer $q$ behavior of
the polarization function (see Eqs.~(\ref{overlapV},
\ref{integralV}) for the explicit calculation)
 \bq\label{Pi(q)gaped}
\Pi(q) = \fr{-2}{3\pi}\fr{q^2}{V} \ .
 \ee
According to Eqs.~(\ref{epsilon}, \ref{epsilonPi}) this results in
the absence of electron's interaction contribution to the
dielectric constant in gapped bilayer graphene at very large
distances.

The fact that at low momentum transfer the $\sim q^2$ behavior of
the polarization function Eq.~(\ref{Pi(q)gaped}) is due to the
small overlap of the spinors $\psi^\dagger_{+}\psi_{-}'\sim q$,
while the nontrivial integration goes over the large-momentum
region, $p\sim \sqrt{t_\perp V}/v_F$, suggests that this result
may be valid also for very small values of $q$, where at $q\sim
V/v_F$ the dispersion $\eps(q)$ is described properly by the full
four-band Hamiltonian Eq.~(\ref{HamK}). Indeed, in order to
calculate the polarization function directly from the four-band
Hamiltonian one would need to add in Eq.~(\ref{Pi(q)}) a summation
over the double set of $+$ and $-$ states, $\psi_{i\pm}$, and to
modify the dispersion $\eps(p)$ in Eq.~(\ref{eps2band})
accordingly. However, also in this case, the product of two
eigenvectors vanishes at small momentum transfer,
$|\psi^\dagger_{i+}\psi_{j-}'|\sim q$, leading to the $\sim q^2$
smallness of $\Pi(q)$. As a fact, the contribution to the
remaining integral from the new terms added into Eq.~(\ref{Pi(q)})
will be suppressed due to the larger denominators ($\sim t_\perp$
instead of $\sim V$). The modification of the term in $\Pi(q)$
describing the transitions between the two lower energy bands of
the four-band model will also lead to corrections of the small
relative order $\sim V/t_\perp$.

The result Eq.~(\ref{Pi(q)gaped}) reveals a nontrivial
distribution of charges induced in bilayer graphene with an
interlayer voltage. First, as we saw in Eqs.~(\ref{screeningTF},
\ref{qTF}), for the case of a very small gap $V$ the negative
charge of an electron is compensated by the positive charge cloud
at distances $\sim 1/q_{\rm TF}$. However, as we see in
Eq.~(\ref{Pi(q)gaped}), the screening of the original electron
charge starts to disappear at distances $\sim\hbar
v_F/\sqrt{t_\perp V}$. This implies the existence of a second very
large negative charge cloud restoring the original electron's
charge.

The low-momentum-transfer formula Eq.~(\ref{Pi(q)gaped}) is valid
for $ q\ll \sqrt{t_\perp V}/\hbar v_F$, when the polarization
operator is small compared to the gapless case, $\Pi(q)\ll
\Pi_{V=0}$. Interestingly, however, this does not imply that this
"small" polarization operator is not sufficient to modify strongly
the electrons' interaction. Indeed, combining Eqs.~(\ref{epsilon},
\ref{epsilonPi}, \ref{Pi(q)gaped}) we write
 \bq\label{Uq2}
U(q)=\fr{2\pi e^2}{\epsilon_0 q +q^2/q_0} \ ,
 \ee
where $q_0=3V/(4e^2)\sim V/(\hbar v_F)$. This formula is still
valid for $q\gg q_0$, where the second, interaction induced term
in the denominator dominates.

The coordinate representation of the interaction Eq.~(\ref{Uq2})
is found via the standard Fourier transformation. The essential
part of the calculation of $U(r)$ reduces then to find the
integral~\cite{GradshteynRyzhik}
 \bq
\int_0^\infty \fr{J_0(k)dk}{x} =\fr{\pi}{2}[H_0(x)-N_0(x)] \ ,
 \ee
where $x=\epsilon_0 q_0 r$ and $J_0, N_0, H_0$ are the Bessel,
Neumann and Struve functions, resp. For $\hbar v_F/\sqrt{t_\perp V}\ll
r\ll \hbar v_F/V$ this gives
 \bq\label{Urlog}
U(r)\approx \fr{3V}{4}\ln \left(\fr{\hbar v_F}{rV}\right) \ .
 \ee
At larger distances, the potential Eq.~(\ref{Urlog}) crosses over
into the usual Coulomb potential $U(r)=e^2/\epsilon_0 r$, while at
shorter distances it transforms into $U(r)\approx e^2/\epsilon_0 r
(q_{\rm TF} r)^2$, which is the screened interaction in bilayer
graphene without a gap~\cite{Hwang08}.

\section{Derivation of Eq.~(\ref{Pi(q)gaped})}

A straightforward calculation of the overlap of the two spinors
in Eqs.~(\ref{psi+}, \ref{psi-}) to leading order in small $q$
gives
 \begin{align}\label{overlapV}
&|\psi_+^\dagger \psi_-'|^2 =\fr{4}{[p^4+(V+\eps)^2]^2}\\
&\times\left[ p^2 q^2(V+\eps)^2 -2\fr{(\vec p\vec
q)^2p^4(V+\eps)}{\eps} +\fr{(\vec p\vec q)^2p^8}{\eps^2}\right] .
\nonumber
 \end{align}
Here, for a moment, we set $\hbar=t_\perp=v_F=1$. Substitution of
this into Eq.~(\ref{Pi(q)}) followed by the angle integration and
changing variables from $p$ to $x=p^4 v_F^4/(t_\perp^2 V^2)$ leads
to
 \bq\label{integralV}
\Pi(q) = \fr{-g}{2\pi}\fr{q^2}{V} \int_0^\infty
\fr{e^2(1+e)^2-xe(1+e)+x^2/2}{e^3[(1+e)^2+x]^2} dx\ ,
 \ee
where $e=\sqrt{1+x}$. Changing the integration variable from $dx$
to $de$ we now find
 \bq
\Pi(q) = \fr{-g}{6\pi}\fr{q^2}{V} \ .
 \ee

%
%
%


\begin{thebibliography}{99}
\bibitem{endnote} In spite of it's simplicity, to the best of
our knowledge, the result that screening~\cite{Hwang08} prevents
the Wigner crystallization in ungapped bilayer graphene was never
claimed in the existing literature. See the discussion after
Eq.~(\ref{screening4}).
\end{thebibliography}
\end{document}